%% file: Vehlhaber.Salazar.ACC24.tex
\documentclass[10pt,conference]{IEEEconf} 

\usepackage{amsmath,amsfonts}
\usepackage{dsfont}
\usepackage{algorithmic}
\usepackage{algorithm}
\usepackage{array}

\usepackage{dsfont}
\usepackage[caption=false,font=normalsize,labelfont=sf,textfont=sf]{subfig}
\usepackage{textcomp}
\usepackage{stfloats}
\usepackage[hyphens]{url}
\usepackage{hyperref}
\usepackage[hyphenbreaks]{breakurl}
\usepackage{verbatim}
\usepackage{graphicx}
\usepackage{cite}
\usepackage{booktabs}
\usepackage{multirow}
\usepackage{comment}
\usepackage{color}
\usepackage{units}
\usepackage{placeins}

\usepackage{amsthm}
\IEEEoverridecommandlockouts

\newif\ifmargincomments 
\margincommentstrue

\newif\ifextendedversion 
\extendedversionfalse

\ifmargincomments

\else

\fi

\newcommand{\vsp}{-7pt} 

\newcounter{probcounter}
\newtheorem{prob}[probcounter]{Problem}

\title{\LARGE \bf Electric Aircraft Assignment, Routing, and Charge Scheduling Considering the Availability of Renewable Energy}

\author{Finn Vehlhaber$^{1}$ and Mauro Salazar$^{1}$
	\thanks{$^{1}$Control System Technology section, Department of Mechanical Engineering,
		Eindhoven University of Technology, 5600 MB  Eindhoven, The Netherlands
		{\tt\small \{f.n.vehlhaber, m.r.u.salazar\}@tue.nl}}%
}
\begin{document}
	\maketitle
	\input{sections/abstract.tex}
	\input{sections/intro.tex}
	\input{sections/methodology.tex}
	\input{sections/results.tex}
	\input{sections/conclusion.tex}
	\FloatBarrier
	\section*{Acknowledgments}
	\noindent
	We thank Emma Flores from NACO for providing us with flight data of the ABC Islands and Dr.\ I.\ New, L.\ Pedroso, M.\ Clemente, J.P.\ Bertucci and O.\ Borsboom for proofreading this paper and their helpful comments.
	This publication is part of the project Green Transport Delta – Elektrificatie with project number MOB21004 of the program R\&D Mobiliteitssectoren regeling, which is (partly) financed by the Dutch Ministry of Economic Affairs and Climate.
	
	\bibliography{../../bibliography/main.bib,../../bibliography/SML_papers.bib}
\end{document}

%% file: sections/abstract.tex
\begin{abstract}
Electric airplanes are expected to take to the skies soon, finding first use cases in small networks within hardly accessible areas, such as island communities.
In this context, the environmental footprint of such airplanes will be strongly determined by the energy sources employed when charging them.
This paper presents a framework to optimize aircraft assignment, routing and charge schedules explicitly accounting for the energy availability at the different airports, which are assumed to be equipped with renewable energy sources and stationary batteries.
Specifically, considering the daily travel demand and weather conditions forecast in advance, we first capture the aircraft operations within a time-expanded directed acyclic graph, and combine it with a dynamic energy model of the individual airports.
Second, aiming at minimizing grid-dependency, we leverage our models to frame the optimal electric aircraft and airport operational problem as a mixed-integer linear program that can be solved with global optimality guarantees.
Finally, we showcase our framework in a real-world case-study considering one week of operations on the Dutch Leeward Antilles.
Our results show that, depending on weather conditions and compared to current schedules, optimizing flights and operations in a renewable-energy-aware manner can reduce grid dependency from 18 to 100\%, whilst significantly shrinking the operational window of the airplanes.
\end{abstract}

%% file: sections/intro.tex
\section{Introduction}
Civil aviation accounts for two percent of global carbon dioxide emissions and demand for flights is steadily growing~\cite{iea2023}. To meet emission reduction targets, manufacturers are working on increasing their aircraft efficiency and investigating the use of alternative fuels such as sustainable aviation fuels, hydrogen, or battery-electric propulsion~\cite{Eurocontrol2022, Hepperle2012}. Meanwhile, policy makers are enacting laws to reduce the dependence on planes, for instance through recent bans on short-haul flights that have the worst carbon footprint~\cite{Sumaila2023}, thus promoting a shift to passenger rail. Yet short-haul aviation is a necessary service in some otherwise inaccessible areas such as island chains or very sparsely populated landscapes. In fact, some countries deem certain flight routes as so essential that they are operated despite being unprofitable. The European Union designates these routes public service obligations (PSOs) and subsidizes airlines that operate them. A similar concept exists in the US with their essential air service (EAS) for remote regions. Such routes may be among the first candidates for electric aviation~\cite{Kinene2022}. 

In recent years, several manufacturers have announced electric airplanes and it will only be a matter of time until they take to the skies. They promise clean and quiet operation but are limited by current battery technology. Unfortunately, current models cannot---or may never---offer ranges or passenger volumes that can rival state-of-the-art passenger airplanes~\cite{Uffelen2018}, which is why their use cases will be limited to short routes with low demand. What is more, an efficient and clean operation requires a ground infrastructure that can supply renewable energy with which to charge these planes~\cite{Cox2023}.
In order to maximize its sustainability, this new mobility paradigm requires particular attention to the weather forecast and flexible scheduling around the availability of renewable energy. 
\begin{figure}[t!]
	\centering
	\includegraphics[width=0.9\columnwidth]{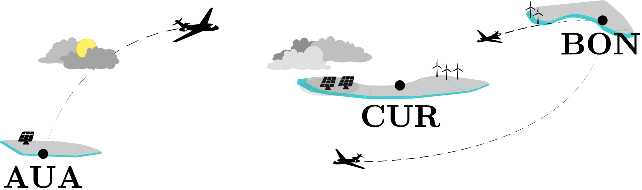}
	\caption{The Dutch Leeward Antilles Aruba, Curaçao, and Bonaire as an example for sustainable regional aviation with the location of their airports labeled with IATA identifiers. \label{fig:ABCislands}}
\end{figure}

Against this background, this paper provides a framework to optimize electric aircraft and airport operations, explicitly accounting for the availabiltiy of renewable energy.
\paragraph*{Related Literature}
This paper is related to three streams of research, namely electric aircraft operations, fleet assignment, and routing.
Despite the absence of electric aircraft in operation, several authors have started to consider its implications for infrastructure and flight operations. Early works explored the sizing of the required charging infrastructure at the airports, under consideration of both battery-swapping and charging, culminating in a framework for obtaining the required number of charging stations for a given electrified flight schedule at a single regional airport~\cite{Bigoni2018,Trainelli2021}. Similarly, Mitici et al.\ showcase a framework for minimizing investment costs of a large airport supporting electric aviation through sizing of the aircraft fleet and charging stations~\cite{Mitici2022}, again considering both battery-swapping and -charging. While the schedule is assumed to be given, the aircraft mission assignment is optimized along with the charging windows.  Framed as a mixed-integer linear program (MILP), the aforementioned works achieve optimality guarantees and good tractability.
Conversely, other authors leverage iterative algorithms from scheduling theory~\cite{Justin2020} or discrete-event simulation (DES)~\cite{Doctor2022} to investigate how electric aircraft affect the operations and infrastructure requirements of an airport, the latter finding that an initial roll-out of electric aviation will have a minor operational influence, even at large airports.
On a broader scale, Kinene et al. design the charging network for regional electric aviation, strategically selecting some airports as charging bases to minimize capital expenditure while increasing connectivity~\cite{Kinene2023}.
While most authors assume the energy to be provided by the national power grid, van Amstel optimizes the energy and charging infrastructure of an airport to only exploit renewable energy sources~\cite{Amstel2023}. In his work, the  flight schedule is given but can be varied slightly in favor of a more energy-efficient operation. Other authors account for the carbon content in the local energy mix and analyze the network to design regional routes and assign the fleet to achieve environmentally friendly operation~\cite{Justin2023}. Whilst the existing literature has explored the infrastructure sizing problem for both single airports and flight networks extensively, there exists a gap for the planning of electric air mobility operations. The case of the desired reliance on only renewable energy is a novelty for electric aircraft scheduling that has yet to be analyzed further.

Fleet assignment and route design have been popular topics in operations research since the advent of commercial aviation~\cite{Ferguson1954,Richardson1976} to improve aircraft utilization, fuel savings, and thus revenue. Historically, due to computational limitations, the schedule of an airline was found via a sequence of these algorithms. Barnhart et al. introduced a string-based model that can achieve fleeting and routing simultaneously and yields feasible schedules directly~\cite{Barnhart1998}. More recently, authors have taken integrated approaches that improve the schedule significantly, for example by accounting for flight speed adjustments that lead to additional fuel savings~\cite{Guerkan2016}. Through techniques like Bender's decomposition, authors even achieve computation times that enable real time updates for schedules that suffer from demand uncertainty~\cite{Cadarso2017}. However, the literature has not yet addressed fleet assignment for electric aircraft.

Roy and Tomlin leverage network flow algorithms for the aircraft routing problem by initializing the network as a time-expanded directed graph~\cite{Roy2007}. After proving that the resulting problem is NP-hard, they introduce sub-optimal solution algorithms that can be efficiently applied to large airspace problems. Another publication describes a framework for the on-demand scheduling and routing of safari planes~\cite{Fuegenschuh2013}, again leveraging a time-expanded digraph. The authors consider capacity and fuel constraints, and propose a time-free relaxation to reduce computational complexity which enables fast solutions for large scale problems with reasonable optimality gaps.
For the vehicle routing problem at large, network flow algorithms have been successful in yielding optimal solutions, even for electric vehicle fleet applications~\cite{SalazarHoushmandEtAl2019,PaparellaHofmanEtAl2023}, also in extended graphs~\cite{RossiIglesiasEtAl2018}. In this context, some authors address the necessity to model the consumption and the battery's state of charge in order to ensure feasible routes, which will apply to electric aircraft routing models as well.

In conclusion, while authors have explored the implications of electric aviation on the airport infrastructure, little work has been done on the planning of electric flight operations. The battery charge requires an additional variable to be tracked, which has been addressed in recent works for eco-routing of electric cars but has yet to be integrated into aircraft routing problems. Additionally, the literature review has highlighted a gap for the consideration of the intrinsic coupling of availability of renewable energy and sustainable electric flight operations.

\paragraph*{Statement of Contributions}
In this paper, we introduce a framework for the scheduling of a fleet of electric airplanes to meet passenger demand while specifically considering the availability of renewable energy at each airport that is part of the network. To this end, we strategically generate the schedule and aircraft routing for a given demand to minimize the grid dependency during the day.

\paragraph*{Organization}
The remainder of this paper is structured as follows: Section \ref{sec:Methodology} outlines a mathematical framework for the integrated fleet assignment and routing for a network of flights relying on renewable energy sources. In Section \ref{sec:Results}, the model is applied to the daily flight operations in the Dutch Caribbean. Finally, Section \ref{sec:Conclusion} draws the conclusions and provides an outlook on implications and possible extensions.

%% file: sections/methodology.tex
\section{Methodology\label{sec:Methodology}}
In this section, we introduce the flight network as a directed acyclic graph (DAG) and outline the energetic models of the aircraft and airport. Thereafter, we introduce the optimization problem.

\subsection{Flight Network}
Here we explain how the graph modeling the flight network is constructed and introduce the necessary terminology. We consider a network of airports $a \in \mathcal{H}$, that is served by a set of aircraft denoted by $\mathcal{P}$. The daily time of operation is discretized into time instances $t \in \mathcal{T}$ with time step $\Delta t$. The set of all possible flight connections $f$ between airports in the network is $\mathcal{F}$, whereby each flight connection takes a number of time steps, $t^f$, which is chosen as the integer closest to the actual time of flight divided by the time step. Each flight connection has a given demand $\mathcal{D}^f$ for the day that needs to be satisfied.

We model the flight network as a DAG $\mathcal{G} = \left(\mathcal{V},\mathcal{A}\right)$. A generic vertex $i \in \mathcal{V}$ corresponds to the space-time tuple $i = (a,t)$, where we denote $i_\mathrm{s} = a$, and $i_\mathrm{t} = t$.

Connecting the vertices is $(i,j) \in \mathcal{A}$, the set of edges, which is constructed as follows:
Between vertices of the same airport that are consecutive in time, there exist $|\mathcal{T}|-1$ ground edges per airport, denoted by the set $\mathcal{A}^\mathrm{g} \subset \mathcal{A}$, where 
\par\nobreak\vspace{\vsp}\begin{small}\begin{equation*}
		 \mathcal{A}^\mathrm{g} = \left\{ (i,j) : j_\mathrm{s} = i_\mathrm{s}, \; j_\mathrm{t} = i_\mathrm{t} + 1\right\} \; ,
\end{equation*}\end{small}%
such that $|\mathcal{A}^\mathrm{g}| = |\mathcal{H}| \cdot (|\mathcal{T}|-1)$.
In addition to the ground edges, we add for each flight a set flight edges $\mathcal{A}^f \subset \mathcal{A}$, where for a flight $f$ from $a$ to $b$ we define the set
\par\nobreak\vspace{\vsp}\begin{small}\begin{equation*}
	\mathcal{A}^f = \left\{(i,j) : i_\mathrm{s} = a, \; j_\mathrm{s} = b, \; j_\mathrm{t} = i_\mathrm{t} + 1  \right\} \quad \forall f \in \mathcal{F} \; ,
\end{equation*}\end{small}%
consequently, with $\mathcal{A}^\mathcal{F} = {\bigcup}_{f \in \mathcal{F}} \mathcal{A}^f $, $\mathcal{A} =  \mathcal{A}^\mathcal{F} \cup \mathcal{A}^\mathrm{g}$.
\begin{figure}[t!]
	\centering
	\includegraphics[width=0.8\columnwidth]{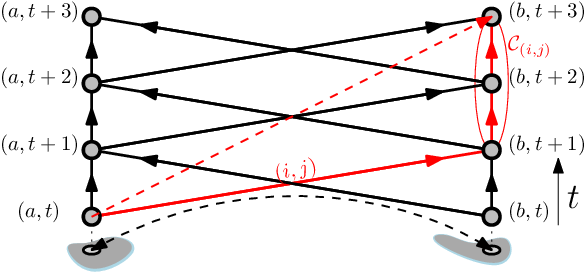}
	\caption{Visualization of the DAG with an example of a flight with a time of flight of 3 time steps (dashed red arrow) and its corresponding flight edge $(i,j)$ and virtual flight edges in the set $\mathcal{C}_{(i,j)}$ in red. \label{fig:DAGvisual}}
\end{figure}
\subsection{Flight Assignment and Aircraft State of Charge}
Each aircraft of the fleet traverses the graph, starting at an airport at time $t$ and terminating at time $|\mathcal{T}|$. We denote an aircraft's origin vertex $o^k \; \forall k \in \mathcal{P}$, and it must be exactly one of the vertices that represent the initial time at either airport. Similarly, the aircraft's destination vertex is $d^k \; \forall k \in \mathcal{P}$, being one of the vertices representing the final time.

We leverage a network flow model formulation where each edge has an associated binary variable $x_{(i,j)}^k\in\{0,1\}$ for each aircraft in the fleet. For a continuous path through time for each aircraft, we require that  
\par\nobreak\vspace{\vsp}\begin{footnotesize}\begin{equation}
		\mathds{1}_{j=o^k} + \sum_{i:(i,j)\in \mathcal{A}} x_{(i,j)}^k = \mathds{1}_{j=d^k} + \sum_{j:(j,l)\in \mathcal{A}} x_{(j,l)}^k \quad \forall j \in \mathcal{V}; \, \forall k \in \mathcal{P} \; , 
		\label{eq:continuity}
\end{equation}\end{footnotesize}%
where $\mathds{1}_{j=o^k}$ and $\mathds{1}_{j=d^k}$ are indicator functions for the origin and destination vertex of each airplane, respectively.

Each flight connection needs to be flown at least as many times as required by the demand, i.e.,
\par\nobreak\vspace{\vsp}\begin{small}\begin{equation}
		\sum_{k \in \mathcal{P}}\sum_{(i,j)\in \mathcal{A}^f} x_{(i,j)}^k \geq \mathcal{D}^f \quad \forall f \in \mathcal{F} \; .
		\label{eq:demand}
\end{equation}\end{small}%
Every flight $f$ has a time of flight  that is some multiple $t^f$ of the time step $\Delta t$. For flights where $t^f>1$ we consider the next $t^f - 1$ ground edges as virtual flight edges that need to be traversed before the plane can enter another flight edge, as exemplified in Fig. \ref{fig:DAGvisual}. To this end, we introduce the auxiliary set
\par\nobreak\vspace{\vsp}\begin{small}\begin{align*}
		\mathcal{C}_{(i,j)} = &\Big\{ (l,m) : \, l_\mathrm{s} = m_\mathrm{s} = j_\mathrm{s} , \, l_\mathrm{t} = j_\mathrm{t} + \tau , \, m_\mathrm{t} = j_\mathrm{t} + \tau + 1 , \\
		&\forall \tau \in \{ 1,\ldots,t^f - 1 \}  \Big\}\subset\mathcal{A}^\mathrm{g} \quad \forall (i,j) \in \mathcal{A}^f, \, \forall f \in \mathcal{F} .
\end{align*}\end{small}%
Then we introduce the constraint
\par\nobreak\vspace{\vsp}\begin{small}\begin{equation}
		x_{(l,m)}^k \geq x_{(i,j)}^k \quad \forall (l,m) \in \mathcal{C}_{(i,j)}, \,\forall (i,j) \in \mathcal{A}^\mathcal{F}, \, \forall k \in \mathcal{P} \; ,
		\label{eq:virtual.flight.edges}
\end{equation}\end{small}%
which forces the path to consist of at least $t_f-1$ ground edges after every flight edge.

For small enough time steps, we must also enforce that the number of aircraft on the same flight edge cannot exceed a certain maximum, as this would have resulted from them starting too closely together. Therefore, we add that
\par\nobreak\vspace{\vsp}\begin{small}\begin{equation}
		\sum_{k \in \mathcal{P}} x_{(i,j)}^k \leq K_\mathrm{max} \quad \forall (i,j) \in \mathcal{A}^\mathcal{F} \; ,
		\label{eq:no.same.start}
\end{equation}\end{small}%
where $K_\mathrm{max}$ is the maximum amount of planes that can start per time step.

When traversing a flight edge, i.e., when flying, the aircraft expends energy $\Delta E^f$ which we calculate as the sum of the energy required for take-off and climb and that for the flight in cruise, the latter of which can be estimated using an adaptation of the Breguet range equation~\cite{Hepperle2012} as
\par\nobreak\vspace{\vsp}\begin{small}\begin{equation}
		\Delta E^f =  \frac{m\cdot g \cdot h_\mathrm{cruise}}{\eta_\mathrm{TO}} + \frac{m \cdot g}{\eta_\mathrm{cruise} \cdot \nicefrac{L}{D}} \cdot d^f \; ,
		\label{eq:flight.energy}
\end{equation}\end{small}%
where $m$ is the mass of the aircraft, $g$ is the gravitational acceleration, $h_\mathrm{cruise}$ the height at which the plane is flying during cruise, $d^f$ the distance of the flight, and $\nicefrac{L}{D}$, $\eta_\mathrm{TO}$, and $\eta_\mathrm{cruise}$ the aerodynamic efficiency of the airplane and the take-off and cruise efficiencies of its powertrain, respectively.

On ground edges that are part of the path, an aircraft can recharge, i.e.,
\par\nobreak\vspace{\vsp}\begin{small}\begin{align}
	P_{\mathrm{c,}(i,j)}^k &\geq 0  \; &\forall (i,j) \in \mathcal{A}^\mathrm{g}\; ,\\
	P_{\mathrm{c,}(i,j)}^k &\leq P_\mathrm{c,max} \; &\forall (i,j) \in \mathcal{A}^\mathrm{g}\; ,\\
	P_{\mathrm{c,}(i,j)}^k &\leq x_{(i,j)}^k \cdot M \; &\forall (i,j) \in \mathcal{A}^\mathrm{g}\; ,
\end{align}\end{small}%
where $P_\mathrm{c,max}$ is the maximum charging power and we leverage the big-M formulation \cite{RichardsHow2005}, with $M$ being a large number.

On ground edges that serve as virtual flight edges after a flight edge, however, it is impossible to recharge as the plane is technically still flying, which is why we add that
\par\nobreak\vspace{\vsp}\begin{small}\begin{equation}
		P_{\mathrm{c,}(l,m)}^k \leq (1 - x_{(i,j)}) \cdot M \quad \forall (l,m) \in \mathcal{C}_{(i,j)} \; .
\end{equation}\end{small}%
For ease of notation, we now refer to all quantities on edges between the same two consecutive time layers by the same $t$. We exemplify this notation with the charging power
\par\nobreak\vspace{\vsp}\begin{small}\begin{equation*}
P_{\mathrm{c,}(i,j)}^k = P_\mathrm{c}^{a,k} (t) \quad \forall \left\{ (i,j) \in \mathcal{A}^\mathrm{g} : i = (a,t)\right\}, \; \forall k \in \mathcal{P} \, .
\end{equation*}\end{small}%
Then, to track the battery state of energy of an aircraft to ensure feasible paths, we define at every time layer $E_\mathrm{p} (t)$ with its dynamics as
\par\nobreak\vspace{\vsp}\begin{small}\begin{equation}
		E_\mathrm{p}^k (t+1) = E_\mathrm{p}^k (t) + \Delta E_\mathrm{p}^k(t) \; ,
\end{equation}\end{small}%
where
\par\nobreak\vspace{\vsp}\begin{small}\begin{equation}
		\Delta E_\mathrm{p}^k(t) = \sum_{a\in \mathcal{H}} P_\mathrm{c}^{a,k} (t) \cdot \Delta t - \sum_{f\in\mathcal{F}}\underset{\tiny \left\{ (i,j)\in \mathcal{A}^f : i_\mathrm{t} = t\right\}}{\sum x_{(i,j)}^k\cdot \Delta E^f } \; .
		\label{eq:delEp}
\end{equation}\end{small}%
Here, the second term collects the energy expenditure on all visited flight edges of this time step. 

We observe that for \eqref{eq:delEp}, we can exploit the structure of the digraph, along with the implications from \eqref{eq:continuity}: For each time step, i.e., between the nodes of two consecutive time layers, exactly one edge lies on the path. Therefore, \eqref{eq:delEp} yields for each time step either the energy recharged at one airport or the energy expended during one flight, but never both.
\subsection{Energy Model of the Airport}
\begin{figure}[t!]
	\centering
	\includegraphics[width=\columnwidth]{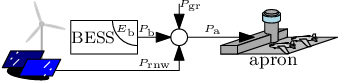}
	\caption{Energetic Model of an airport with renewable energy sources and stationary battery energy storage system (BESS). Arrows indicate direction of positive power flow. \label{fig:AirportEnergyModel}}
\end{figure}

In this section, we explain how the aircraft traversing the flight network are connected to their respective airports via local dynamic energy models.
As shown in Fig.~\ref{fig:AirportEnergyModel}, we assume each airport to be equipped with a battery energy storage system (BESS) that can store and supply power, a connection to the grid, a connection to local renewable power sources, and an apron where the gates and chargers for the electric airplanes are located. 

The power drawn at the apron during one time step is obtained from the sum of all chargers, i.e.,
\par\nobreak\vspace{\vsp}\begin{footnotesize}\begin{equation}
		P_\mathrm{a}^a (t)= \sum_{k \in \mathcal{P}} P_{\mathrm{c}}^{a,k} (t) \quad \forall t \in \mathcal{T},\, \forall a \in \mathcal{H},
\end{equation}\end{footnotesize}%
This power is potentially subject to limits, imposed by
\par\nobreak\vspace{\vsp}\begin{small}\begin{equation}
		P_\mathrm{a}^a (t) \in [0,P_\mathrm{a,max}] \quad \forall t \in \mathcal{T},\, \forall a \in \mathcal{H} \; ,
\end{equation}\end{small}%
where $P_\mathrm{a,max}$ is the maximum power that can be supplied to the apron at the airport.

For safety reasons, planes can only start, land, and charge during the time of operations of the airport, previously denoted by $\mathcal{T}$. The renewable power sources, however, can supply power during the whole day, and the BESS can store said power to reduce curtailment. Therefore, we distinguish between the set of time instances for the whole day, $\mathcal{T}_\mathrm{d}$, and that for the time of operations, $\mathcal{T} \subset \mathcal{T}_\mathrm{d}$. Consequently,
\par\nobreak\vspace{\vsp}\begin{small}\begin{equation}
		P_\mathrm{a}^a (t) = 0 \quad \forall t \notin \mathcal{T} \; .
\end{equation}\end{small}%
We express the power split at the airport as
\par\nobreak\vspace{\vsp}\begin{small}\begin{equation}
		P_\mathrm{gr}^a (t) = P_\mathrm{a}^a (t) + P_\mathrm{aux} - P_\mathrm{rnw}^a (t) - P_\mathrm{b}^a(t) \; ,
\end{equation}\end{small}%
where $P_\mathrm{gr}^a (t)$, $P_\mathrm{rnw}^a (t)$, and $P_\mathrm{b}^a (t)$ are the powers drawn from the grid and supplied from renewables and the BESS, respectively, and $P_\mathrm{aux}$ is the power draw from all auxiliary systems of the airport, here assumed constant.
We model the dynamics of the BESS as
\par\nobreak\vspace{\vsp}\begin{small}\begin{align}
		E_{\mathrm{b}}^a (t+1) &\leq E_{\mathrm{b}}^a (t) - \eta_\mathrm{b} \cdot P_\mathrm{b}^a (t) \cdot \Delta t \quad &\forall t,t+1 \in \mathcal{T}_\mathrm{d} \; , \\
		E_{\mathrm{b}}^a (t+1) &\leq E_{\mathrm{b}}^a (t) - \frac{1}{\eta_\mathrm{b}} P_\mathrm{b}^a (t) \cdot \Delta t \quad &\forall t,t+1 \in \mathcal{T}_\mathrm{d} \; ,
\end{align}\end{small}%
where $\eta_\mathrm{b}$ is an efficiency term for charging and discharging the battery.

Periodicity constraints ensure that the battery ends its day with the same state of energy it started with, i.e.,
\par\nobreak\vspace{\vsp}\begin{small}\begin{equation}
		E_{\mathrm{b}}^a (t_0) = E_{\mathrm{b}}^a (t_\mathrm{f}) \quad \forall a \in \mathcal{H} \; ,
\end{equation}\end{small}%
where $t_0$ and $t_\mathrm{f}$ are the time steps at the start and the end of the day, respectively, and the battery energy is subject to the limits
\par\nobreak\vspace{\vsp}\begin{small}\begin{equation}
		E_{\mathrm{b}}^a (t) \in [E_{\mathrm{b,min}}^a, E_{\mathrm{b,max}}^a] \quad \forall t \in \mathcal{T}_\mathrm{d}, \, \forall a \in \mathcal{H} \; .
\end{equation}\end{small}%
At the beginning of operations, we require the BESS to have a state of charge (SoC) of at least $\xi_\mathrm{b,init}$ in order to facilitate the availability of energy for unforeseen circumstances:
\par\nobreak\vspace{\vsp}\begin{small}\begin{equation}
		E_{\mathrm{b}}^h (t_\mathrm{0,ops}) \geq \xi_\mathrm{b,init} \cdot E_{\mathrm{b,max}}^a \quad \forall a \in \mathcal{H} \; .
\end{equation}\end{small}%
The battery power is limited through
\par\nobreak\vspace{\vsp}\begin{small}\begin{equation}
		P_{\mathrm{b}}^a (t) \in [P_{\mathrm{b,min}}^a, P_{\mathrm{b,max}}^a] \quad \forall t,t+1 \in \mathcal{T}_\mathrm{d} \; .
		\label{eq:battery.limits}
\end{equation}\end{small}%
Given the solar irradiation $I_\mathrm{s}^a (t)$, we find the solar energy obtained through the solar cells at each airport from
\par\nobreak\vspace{\vsp}\begin{small}\begin{equation}
		P_{\mathrm{rnw}}^a (t) \leq I_\mathrm{s}^a (t) \cdot A_\mathrm{sc}^a \cdot \eta_\mathrm{sc} \quad\forall t,t+1 \in \mathcal{T}_\mathrm{d}, \, \forall a \in \mathcal{H} \; ,
		\label{eq:renewable.power}
\end{equation}\end{small}%
where $A_\mathrm{sc}^a$ is the area of the solar cells installed at that airport and $\eta_\mathrm{sc}$ is the conversion efficiency of the cells. For the sake of simplicity we do not include other renewable energy sources such as wind or water power here, which can, however, readily be included within our framework.

For this paper we assume the airport to be a consumer only, i.e.,
\par\nobreak\vspace{\vsp}\begin{small}\begin{equation}
		P_{\mathrm{gr}}^a (t) \geq 0 \quad\forall t,t+1 \in \mathcal{T}_\mathrm{d}, \, \forall a \in \mathcal{H} \; ,
		\label{eq:grid.constraint}
\end{equation}\end{small}%
but with financial objective functions this could be changed to have the airport acting as a producer as well.

\subsection{Objective}
As remote communities often suffer from a limited grid connection, we aim at finding a flight schedule that reduces the energy drawn from the grid with a given fleet size. To this end, we frame the optimization problem as follows.
\begin{prob}[Grid energy minimization]\label{prob:Egr}The flight schedule that meets the required demand and best exploits the availability of renewable energy is obtained through
	\begin{align*}
		\min \; & E_\mathrm{gr} = \sum_{a \in \mathcal{H}}\sum_{t \in \mathcal{T}_\mathrm{d}} P_{\mathrm{gr}}^a (t) \cdot \Delta t\\
		\mathrm{s.t.} \;& \eqref{eq:continuity} - \eqref{eq:grid.constraint}.
	\end{align*}
\end{prob}
With minor adjustments, the introduced framework can also lend itself to optimization with other objectives and variables in mind, such as fleet or infrastructure sizing, which are here omitted due to space restrictions.

%% file: sections/results.tex
\section{Results\label{sec:Results}}
In this section, we showcase the effectiveness of the framework in a case study for a flight network in the Dutch Leeward Antilles (ABC islands) and comment on it. 

\begin{table}[t!]
	\centering
	\caption{Number of scheduled flights for a week of operations on the ABC Islands. \label{tab:flight.demand}}
	\begin{tabular}{|l|l|ccccccc|}
		\toprule
		\multirow{2}{*}{\textbf{Origin}} & \multirow{2}{*}{\textbf{Destination}} & \multicolumn{7}{c|}{\textbf{\# Flights}}\\
		~& ~& M&T&W&T&F&S&S\\
		\midrule \midrule
		AUA & CUR&8&5&5&8&8&6&5\\ 
		CUR & AUA&8&5&5&8&8&6&5\\
		BON & CUR&11&10&9&10&11&9&8\\ 
		CUR & BON&11&10&9&10&11&9&8\\
		\bottomrule
	\end{tabular}
\end{table}
The ABC islands are an island community in the Caribbean with sunny weather conditions year-round, making them a perfect candidate for sustainable regional aviation. Inter-island flights are currently conducted with a fleet of BN-2 Islanders and DHC6 Twin Otters operated by the local airline Divi Divi Air~\cite{flightradar24}, which can be replaced with early electric aircraft models in the near future.

We extract all inter-island flights from the daily flight schedules of August\footnote{supplied by NACO International Aviation Consultancy} that are flown by the aforementioned aircraft and assume that the demand can be covered with an initial fleet of eight electric nine-seaters. For the airplanes, we assume a homogeneous fleet of the model Alice, a nine-seater announced by the manufacturer Eviation \cite{Eviation}. The required number of flights per pairing is shown in Table \ref{tab:flight.demand}. At the beginning and end of operations all aircraft are assumed to be parked at the airport in Cura\c{c}ao (CUR), which is the base of the airline. Along with the flight demand, we use the weather conditions of that same day, where the solar irradiation at each airport was obtained from publicly available weather data \cite{Wunderground}. The airports are not equipped with solar panels or a BESS yet, so we assume a solar array of \unit[2000]{m$^2$} and a BESS with \unit[1]{MWh} on each.
\begin{figure}[t!]
	\centering
	\includegraphics[width=0.8\columnwidth]{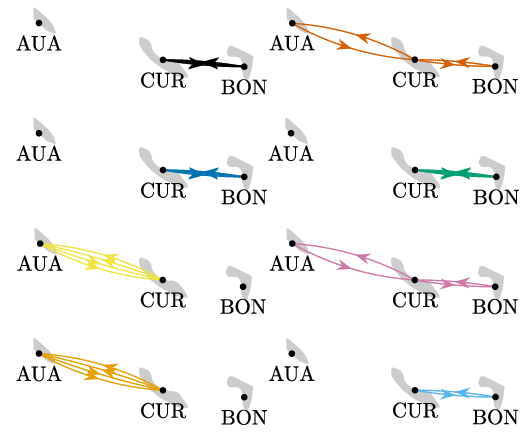}
	\includegraphics[width=\columnwidth]{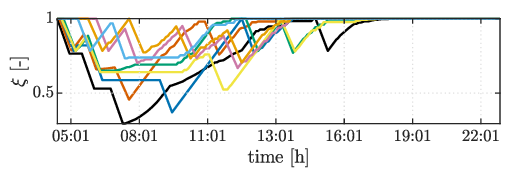}
	\caption{Optimal routes and state of charge evolution of all planes of the fleet serving the flight demand in Table \ref{tab:flight.demand} on the ABC islands with weather conditions of Saturday, August 19. \label{fig:ABC.planes}}
\end{figure}

\begin{figure}[t!]
	\centering
	\includegraphics[width=\columnwidth]{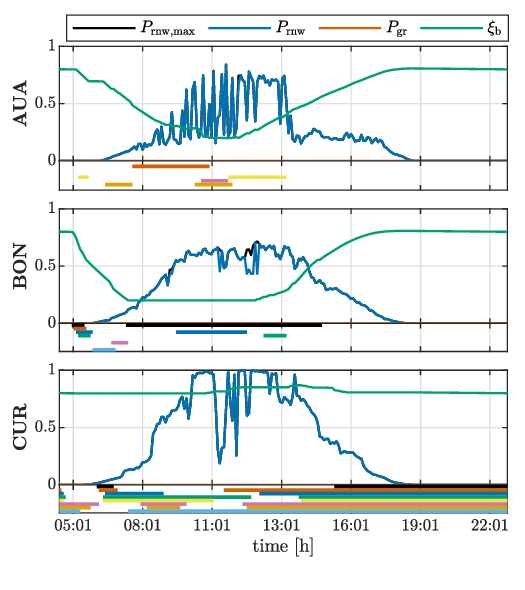}
	\caption{Airport optimal power breakdown and state of charge evolution of the BESS for the routes shown in Fig. \ref{fig:ABC.planes}. Thick bars under each plot mark an aircraft being on the tarmac during that time. \label{fig:ABC.airports}}
\end{figure}

The graph creation and pre-processing is carried out in MATLAB in negligible time. Problem~\ref{prob:Egr} is a MILP that we parse with YALMIP and solve with Gurobi \cite{Loefberg2004,GurobiOptimization2021}. On commodity hardware and with a discretization of $\Delta t =$ \unit[10]{min} we obtain a globally optimal solution within minutes. The results are depicted in Figures~\ref{fig:ABC.planes} and~\ref{fig:ABC.airports}.
\begin{figure}[h!]
	\centering
	\includegraphics[width=\columnwidth]{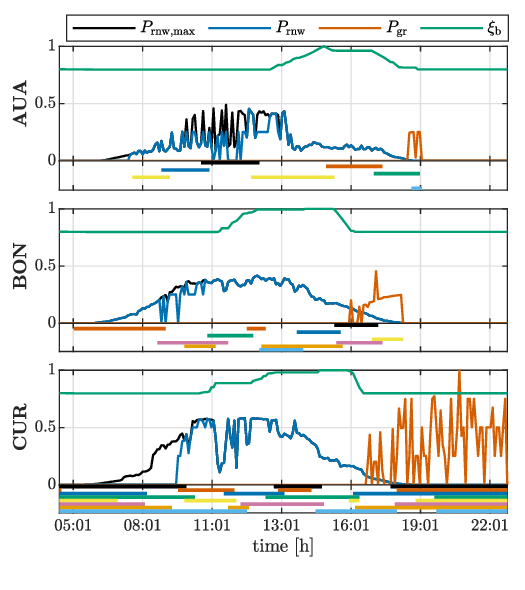}
	\caption{Airport optimal power breakdown and state of charge evolution of the BESS for scheduled operations on Saturday, Aug. 19. Thick bars under each plot mark an aircraft being on the tarmac during that time. \label{fig:ABC.schedule.airports}}
\end{figure}

\begin{figure}[h!]
	\centering
	\includegraphics[width=\columnwidth]{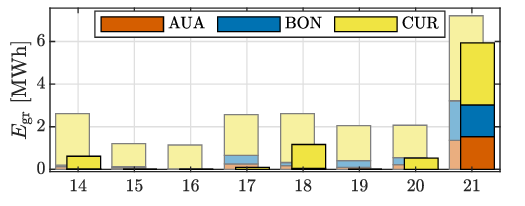}
	\caption{Comparison of grid energy requirements for the scheduled operation (background, muted) and the solar-optimized operation (foreground) for each day from Monday, Aug. 14, to Monday, Aug. 21. Differences in energy requirements between days with the same demand are due to the different weather conditions. \label{fig:ABC.barChart}}
\end{figure}

We compare our optimized schedule to the actual schedule of the same day, whereby we still optimize aircraft charging schedules as shown in Fig.~\ref{fig:ABC.schedule.airports}.
Over the course of the week, Fig.~\ref{fig:ABC.barChart} shows that, depending on the weather conditions, the solar-optimized schedule requires between \unit[18]{\%} and \unit[100]{\%} less energy from the grid than the actual schedule. Moreover, the majority if not all of the total grid energy requirements is at Cura\c{c}ao, where the airplanes start and end their day and which is either origin or destination of all demanded flights, indicating the need for additional design considerations, especially to ensure grid independence even in extreme cases.

%% file: sections/conclusion.tex
\section{Conclusion\label{sec:Conclusion}}
This paper introduced an optimization model for the routing and charge scheduling of electric aircraft to meet a daily passenger demand that reduces grid dependence by explicitly accounting for the availability of renewable energy.
Framed as a mixed-integer linear program, our problem can be efficiently solved with global optimality guarantees.
The framework was applied on a regional aviation network on the Dutch ABC Islands, for which demand and weather conditions were known.
The results showed that a schedule that specifically accounts for solar power supply can reduce grid dependency by up to \unit[100]{\%}, which is of special interest for remote airfields that serve otherwise badly connected communities.

This work can be built upon as follows: First, passenger demand models and weather uncertainty could be introduced to study the robustness of the resulting schedules. 
Second, it would be of interest to explore additional use cases to investigate investment strategies to enhance electric aviation.